\documentclass[twocolumn]{aastex631}

\usepackage{xcolor}
\usepackage{comment}

\newcommand{\teff}{$T_{\!\mbox{\scriptsize\it eff}}$}	% equal to tefftext
\newcommand{\logg}{log\,$g$}
\newcommand{\loggf}{log\,$g_{\!\mbox{\tiny\it F}}$}

\newcommand{\mbol}{$m\,_{\!\mbox{\tiny\scriptsize bol}}$}
\newcommand{\Mbol}{$M_{bol}$}
\newcommand{\zsun}{$Z_\odot$}
\newcommand{\msun}{$M_\odot$}

\newcommand{\hii}{H\,{\sc ii}\rm}

\newcommand{\hei}{He\,{\sc i}\rm}
\newcommand{\heii}{He\,{\sc ii}\rm}

\newcommand{\oiii}{[O\,{\sc iii}]}

\newcommand{\oii}{[O\,{\sc ii}]}

\newcommand{\siliconii}{Si\,{\sc ii}}
\newcommand{\siliconiii}{Si\,{\sc iii}}

\newcommand{\caii}{Ca\,{\sc ii}}
\newcommand{\magnesiumii}{Mg\,{\sc ii}}
\newcommand{\heliumi}{He\,{\sc i}}
\newcommand{\heliumii}{He\,{\sc ii}}

\newcommand{\eg}{{e.g.}}
\newcommand{\ie}{{i.e.}}

\newcommand{\trgb}{{\sc trgb}}
\newcommand{\bsg}{\mbox{\small BSG}}

\newcommand{\hgamma}{H$\gamma$}
\newcommand{\ewhgamma}{$W_{\gamma}$}

\newcommand{\hbeta}{H$\beta$}

\newcommand{\lin}{$\,\lambda$}
\newcommand{\llin}{$\,\lambda\lambda$}

\newcommand{\oh}{12\,+\,log(O/H)}

\renewcommand{\eg}{\mbox{e.g.}}
\renewcommand{\ie}{\mbox{i.e.}}

\newcommand{\fwhm}{\mbox{\small FWHM}}
\newcommand{\mzr}{\mbox{\small MZR}}

\newcommand{\fglr}{\mbox{\sc fglr}}

\newcommand{\leoa}{\mbox{Leo~A}} 

\newcommand{\newhash}{%
	{\settoheight{\dimen0}{C}\kern-.05em \resizebox{!}{\dimen0}{\raisebox{\depth}{\#}}}}

\shorttitle{Blue supergiants in Leo A}
\shortauthors{Urbaneja et al.}

\begin{document}

\title{The metallicity and distance of Leo A from blue supergiants}

%%
%% Use \email to set provide email addresses. Each \email will appear on its
%% own line so you can put multiple email address in one \email call. A new
%% \correspondingauthor command is available in V6.31 to identify the
%% corresponding author of the manuscript. It is the author's responsibility
%% to make sure this name is also in the author list.
%%

%%

\correspondingauthor{Miguel.Urbaneja-Perez}
\email{Miguel.Urbaneja-Perez@uibk.ac.at, bresolin@ifa.hawaii.edu, kud@ifa.hawaii.edu}

\author[0000-0002-9424-0501]{Miguel A. Urbaneja}
\affil{Universit\"at Innsbruck, Institut f\"ur Astro- und Teilchenphysik\\ 
	Technikerstr. 25/8, 6020 Innsbruck, Austria}

\author[0000-0002-5068-9833]{Fabio Bresolin}
\affiliation{Institute for Astronomy, University of Hawaii \\
	2680 Woodlawn Drive \\
	Honolulu, HI 96822, USA}

\author{Rolf-Peter Kudritzki}
\affiliation{Institute for Astronomy, University of Hawaii \\
	2680 Woodlawn Drive \\
	Honolulu, HI 96822, USA}
\affiliation{University Observatory Munich, Scheinerstr. 1, D-81679 Munich, Germany}

%==================================================================================
\begin{abstract}
We have obtained high-quality spectra of blue supergiant candidates in the dwarf irregular galaxy \leoa\ with the Low Resolution Imaging Spectrometer at the Keck~I telescope. From the quantitative analysis of seven B8-A0 stars we derive a mean metallicity $[Z] = -1.35\pm0.08$, in excellent agreement with the gas-phase chemical abundance. From the stellar parameters and the flux-weighted--luminosity relation (\fglr) we derive a spectroscopic distance modulus $m-M = 24.77 \pm 0.11$ mag, significantly larger ($\sim$0.4 mag) than the value indicated by RR Lyrae and other stellar indicators. We explain the bulk of this discrepancy with blue loop stellar evolution at very low metallicity and show that the combination of metallicity effects and blue loop evolution amounts, in the case of \leoa, to a $\sim$0.35~mag
offset of the \fglr\ to fainter bolometric luminosities. We identify one outlier of low bolometric magnitude as a post-AGB star. Its metallicity is consistent with that of the young population, confirming the slow chemical enrichment of \leoa.

\end{abstract}

%==================================================================================


\section{Introduction} \label{sec:intro}
The dwarf irregular galaxy \leoa\ (DDO~69, UGC~5364), at a distance of 0.76~Mpc (\citealt{Nagarajan:2022}),
is among the gas-rich systems with the lowest gas-phase metallicity known in the nearby universe, \oh\,$\simeq$\,7.4\footnote{This value corresponds to [O/H]\,$\simeq$\,$-1.3$ (5\% of the oxygen abundance of the Sun) with the solar oxygen abundance of \citet{Asplund:2009} adopted here.} (\citealt{Ruiz-Escobedo:2018}). 
Its delayed, low-level star formation activity (\citealt{Cole:2007}) has been related to its isolation (\citealt{Cole:2014, Gallart:2015}), as \leoa\ assembled in a low-density environment near the edge of the Local Group, several hundreds of kpc from both the Milky Way and M31 (\citealt{McConnachie:2012}). 
\leoa\ offers the opportunity for a rare glimpse into the properties and chemistry of young stars in unevolved star-forming environments attainable from high-quality spectra.

The first spectroscopic investigation of individual stars in \leoa\ was carried out by \citet{Brown:2007}, who derived the stellar velocity dispersion of the galaxy from spectra of 10 B supergiants (two of these were included in their previous search for hypervelocity stars: \citealt{Brown:2006}) and two \hii\ regions. The spectroscopic analysis of the older population of red giants has been presented by \citet{Kirby:2017}, enlarging the sample of \citet{Kirby:2013}. They derived a mean [Fe/H]\,=\,$-1.67$, only slightly below the gas-phase metallicity, suggesting a slow chemical enrichment in \leoa, in agreement with the flat age-metallicity relation derived from Hubble Space Telescope (HST) photometry by \citet{Cole:2007}.
More recently, \citet{Gull:2022} derived stellar parameters for a sample of massive stars (late O/early B main-sequence objects) in \leoa\ from both HST photometry and ground-based spectroscopy. 

With this paper we contribute to the spectroscopic investigation of individual stars in \leoa\ and the present-day metal content of this galaxy by focusing on blue supergiants (\bsg s), \ie\ evolved massive stars. We use a comprehensive grid of model atmosphere spectra based on detailed and extensive non-LTE calculations to determine stellar temperatures, gravities, luminosities and metallicities together with interstellar extinction. As we have shown in the work by our team over the last two decades (see \citealt{Kudritzki:2008, Kudritzki:2012, Kudritzki:2016, Urbaneja:2008, Urbaneja:2017, Bresolin:2016, Bresolin:2022}, and further references therein) by studying more than a dozen nearby galaxies, this method of quantitative spectral analysis provides accurate information about galaxy metallicity and metallicity gradients as well as distances. It provides an excellent basis to understand galaxy evolution, for instance, by investigating the observed mass-metallicity relationship of galaxies \citep{Kudritzki:2021}. In this regard \leoa, as a potentially extremely metal-poor galaxy, is especially interesting.

The challenge of a \bsg~study in \leoa\ is the relative faintness of the targets ($V$\,$\sim$\,19-21), together with their low metallicity and the correspondingly weak metal lines. This challenge is met with high-quality spectra of high signal-to-noise ratio (up to $\sim$\,200) obtained in a long sequence of exposures with the LRIS spectrograph at the Keck telescope. Our approach to deriving stellar parameters and metallicities is the same we adopted in previous work on galaxies of higher chemical abundances (for a recent example, see \citealt{Bresolin:2022} and references therein).
In Sect.~\ref{sec:observations} we present observations, data reduction and spectral classification of the targets. The quantitative analysis of seven late-B/early-A stars is described in Sect.~\ref{sec:quantitative}. We then use the parameters for these objects to briefly discuss the location of \leoa\ in the stellar mass--metallicity relation of star-forming galaxies (Sect.~\ref{sec:MZR}) and the spectroscopic distance we derive (Sect.~\ref{sec:distance}). Considerations on a post-AGB star that we have serendipitously discovered in \leoa\ are presented in Sect.~\ref{sec:postagb}.
We summarize our results in Sect.~\ref{sec:summary}.

%==================================================================================
%\floattable
\begin{deluxetable*}{lCCcccc}
	\tabletypesize{\footnotesize}	
	\tablecolumns{7}
	\tablecaption{Properties of the spectroscopic targets.  \label{table:1}}
	
	\tablehead{
		\colhead{ID}	     		&
		\colhead{R.A.}	 			&
		\colhead{Decl.}	 			&
		\colhead{$V$}				&
		\colhead{$B-V$}				&
        \colhead{$V-I$}				&
		\colhead{Spectral}\\[-2.5ex]
		\colhead{}       			&
		\colhead{(J2000.0)}       	&
		\colhead{(J2000.0)}       	&
		\colhead{(mag)}            	& 	
        \colhead{(mag)}            	& 
		\colhead{(mag)}             & 															
		\colhead{type} \\[-5ex] } 	
	
	\colnumbers
	\startdata
	\\[-4.5ex]
	01  &   09\; 59\; 16.08  &   30\; 44\; 25.2  &  21.36  & $-$0.19  & $-$0.14  &   (B5 II) \\[-0.4ex]  %   b67
02  &   09\; 59\; 16.94  &   30\; 43\; 48.2  &  19.27  & $-$0.11  & \phantom{$-$}0.00  &   B2 Iab \\[-0.4ex]  %   AC1
03 $\ast$  &   09\; 59\; 17.82  &   30\; 45\; 34.0  &  21.15  & $-$0.03  & \phantom{$-$}0.05 &   (A0 II) \\[-0.4ex]  %   b60
04 $\ast$  &   09\; 59\; 20.21  &   30\; 43\; 52.8  &  19.65  & $-$0.07  & $-$0.01 &   B8 Ib \\[-0.4ex]  %   b49
05 $\ast$  &   09\; 59\; 23.21  &   30\; 45\; 06.3  &  20.27  & $-$0.07  & \phantom{$-$}0.00 &  B8 Ib \\[-0.4ex]  %   b51
06 $\ast$  &   09\; 59\; 25.43  &   30\; 44\; 57.8  &  19.85  & $-$0.06  & $-$0.04 &  B8 Ib \\[-0.4ex]  %   a04
07 $\ast$  &   09\; 59\; 26.34  &   30\; 45\; 26.2  &  19.33  & $-$0.06  & $-$0.02 &  B8 Ib \\[-0.4ex]  %   a01
08  &   09\; 59\; 27.52  &   30\; 44\; 57.9  &  20.31  & $-$0.24  & $-$0.25 &  O9 V \\[-0.4ex]  %   a09
09  &   09\; 59\; 29.72  &   30\; 45\; 05.8  &  21.49  & \phantom{$-$}0.04  & \phantom{$-$}0.09 &  (A0 III) \\[-0.4ex]  %   a40
10 $\ast$  &   09\; 59\; 30.48  &   30\; 44\; 35.2  &  19.70  & \phantom{$-$}0.08  & \phantom{$-$}0.20 &  A0 II \\[-0.4ex]  %   a02
11 $\ast$  &   09\; 59\; 31.67  &   30\; 44\; 43.4  &  19.97  & $-$0.20  &  \phantom{$-$}0.03 & B8 Ib \\[-0.4ex]  %   a07
12  &   09\; 59\; 33.32  &   30\; 44\; 39.2  &  19.80  & $-$0.13  & \phantom{$-$}0.03 &  O7/O8 V \\[-0.4ex]  %   H2
	\\[-2.5ex]
	\enddata
	\tablecomments{Stars identified with the asterisk in column (1) are those analyzed in Sect.~\ref{sec:quantitative}. Photometric information taken from \citet{Stonkute:2014}. Spectral types within parentheses in column (7) are considered uncertain.}
\end{deluxetable*}

%==================================================================================

%==================================================================================
\section{Observations and data reduction} \label{sec:observations}
\subsection{Target selection}\label{subsec:selection}
We selected blue supergiant candidates for the spectroscopic observations from stellar photometry of archival 
HST/ACS imaging (Program: 10590; PI: Cole) in the F475W and F814W filters that we carried out using DOLPHOT (\citealt{Dolphin:2000, Dolphin:2016}). Additional targets were selected outside the ACS field of view from published photometry of bright stars in \leoa\ from the Sloan Digital Sky Survey (SDSS, \citealt{Adelman-McCarthy:2007}). Our final list 
for the multi-object spectroscopy consisted of 12 stars, selected to be brighter than $B=22.0$ and bluer than $B-I=0.1$. These criteria serve the purpose of isolating early-type supergiant candidates of sufficient brightness for our quantitative analysis, affected by only a small amount of reddening.

\begin{figure*}
	\center \includegraphics[width=1\textwidth]{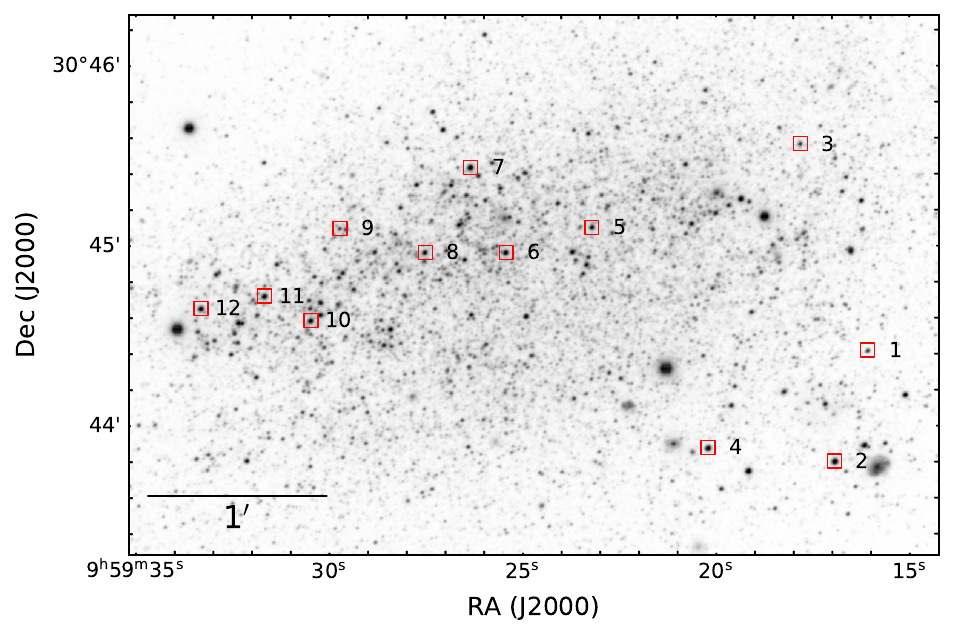}\medskip
	\caption{Identification of the spectroscopic targets in a Subaru $B$-band image of \leoa\ taken from \citet{Stonkute:2014}.}\label{fig:targets}
\end{figure*}

A summary of positions, $V$ magnitudes and $B-V$, \mbox{$V-I$} color indices of the 12 targets is presented in Table~\ref{table:1}. 
The photometric information is extracted from the work of \citet{Stonkute:2014}, which we prefer to adopt here over our original photometry because of the additional color information and the larger spatial coverage. Figure~\ref{fig:targets} displays the location of these objects. 

% - - - - - - - - - - - - - - - - - - - - - - - - - - - - - - - - - - - - - - - - - 
% - - - - - - - - - - - - - - - - - - - - - - - - - - - - - - - - - - - - - - - - - 

% - - - - - - - - - - - - - - - - - - - - - - - - - - - - - - - - - - - - - - - - - 
\begin{figure*}
	\center \includegraphics[width=0.9\textwidth]{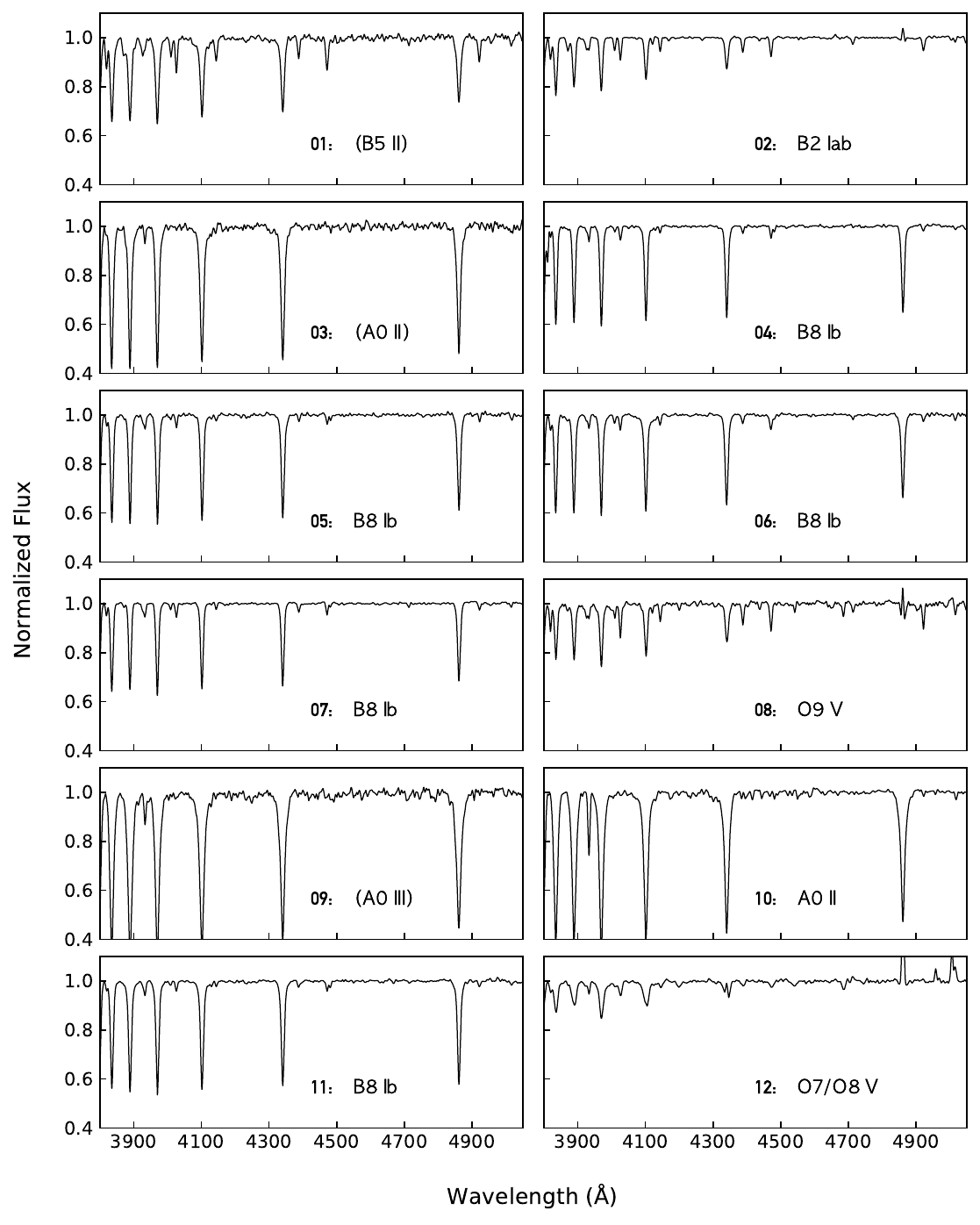}\medskip
	\caption{Compilation of the spectra of the 12 targets observed.}\label{fig:spectra}
\end{figure*}
% - - - - - - - - - - - - - - - - - - - - - - - - - - - - - - - - - - - - - - - - - 

%==================================================================================
\subsection{Spectroscopy}
We collected the spectra of our \bsg\ candidates with the Low Resolution Imaging Spectrometer (LRIS, \citealt{Oke:1995}) at the Keck~I telescope, during the course
of six nights, distributed over three distinct observing campaigns in 2012, 2013 and 2015. \leoa\ was observed at the start of each night, when our main target, NGC~4258, was too low above the horizon to be observed (Kudritzki et al., in preparation).

The multi-object spectroscopy was obtained employing a single slit mask.
With the selected 600/4000 grism and 1\farcs2 slit width our data 
cover the approximate wavelength range 3300--5600~\AA\ with a 
\fwhm\ spectral resolution of about 5\,\AA.
Individual exposure times varied between 1500\,s and 2700\,s, for a total integration time of 5.9\,h.

The data reduction was carried out with {\sc iraf}\footnote{IRAF was distributed by the National Optical Astronomy Observatory, which was managed by the Association of Universities for Research in Astronomy (AURA) under a cooperative agreement with the National Science Foundation.}, and included the usual steps of bias subtraction, flat field correction and wavelength calibration. The spectral extractions for individual exposures were registered with the aid of sky lines and later combined. Finally, we rectified the combined spectra using low-order polynomials, and applied a radial velocity correction based on stellar lines in order to work with spectra in the rest frame.
After completion of these steps, the average signal-to-noise ratio of the coadded spectra ranges between 40 and 220 (median: 140). We display in Figure~\ref{fig:spectra} the final spectra of the 12 targets in the 3800-5050~\AA\ interval.

%==================================================================================
\subsection{Spectral classification} \label{sec:classification}
The classification of early-type stars of low metallicity requires an adjustment to the MK criteria since these are established from Galactic standards. Attempts to devise a stellar classification system for LMC and SMC stars were introduced by \citet{Fitzpatrick:1991} and \citet{Lennon:1997}, respectively. In our previous blue supergiant work in low-metallicity galactic environments (WLM -- [O/H]\,=\,$-0.86$:
%\footnote{We adopt \ohsun 8.69 from \citealt{Asplund:2009}.}
\citealt{Bresolin:2006a}; NGC~3109 -- [O/H]\,=\,$-0.93$: \citealt{Evans:2007}; IC~1613 -- [O/H]\,=\,$-0.79$: \citealt{Bresolin:2007a}) we have followed the suggestion by \citet{Lennon:1997} to base the classification of B-type stars on trends of metal line strengths. For A-type stars, we followed instead \citet{Evans:2003} and \citet{Evans:2004}, while for O stars
we used the work by \citet{Walborn:1971}, \citet{Walborn:1990} and \citet{Walborn:2000}.
In the case of \leoa, where stars have even weaker metal lines, we have relied on the same criteria, including the \caii~K/(\caii~H + H$\epsilon$) line ratio (A stars), the relative strengths of the \heliumi, \siliconii, \siliconiii\ and \magnesiumii\ lines (B stars), and the relative strengths of the \heliumii\ and \heliumi\ lines (O stars).

The equivalent width of the \hgamma\ line, \ewhgamma, was used as a guidance to determine the luminosity classes, following \citet{Balona:1974} and \citet{Azzopardi:1987}. 
Our results show that, in addition to two late-O dwarfs, our sample includes several class II/III B- and A-type objects. The brightest stars are of class Iab/Ib, and range in absolute magnitude between $M_V\simeq-5.1$ to $M_V\simeq-4.1$, significantly fainter than stars we have analyzed in more massive and more actively star-forming galaxies, where $M_V$ can be as bright as $-8$ to $-9$. 

We note that six of our 12 targets were included in the velocity dispersion study of \citet{Brown:2007}. %who concluded that the stellar objects they analyzed were likely B supergiants of class I or II. We confirm that this is indeed the case for the four B stars we have in common
Using spectra of inferior quality compared to our Keck data, \citet{Brown:2007} concluded that 10 of those were likely B supergiants of class I or II, whilst the remaining two objects were \hii\ regions. Of the six objects in common with the aforementioned work, we confirm that four of those that Brown et al. identified as B supergiants are indeed supergiant stars (02, 04, 05 and 07 in Table~\ref{table:1}), whilst the other two, identified as \hii\ regions, are in fact late-O main sequence stars (08 and 12 in Table~\ref{table:1}).

%(02, 04, 05 and 07 in Table~\ref{table:1}), while the remaining two objects identified as \hii\ regions by \citet{Brown:2007} from spectra of inferior quality compared to our Keck data correspond to 
%late-O main sequence stars (08 and 12 in in Table~\ref{table:1}).

Star 08 is the only object we have in common with the work by \citet[their star K1/B1]{Gull:2022}, who classified all of their stars as main-sequence (class V) objects.
Our O9\,V classification matches the one provided by these authors.
We note that, like these authors, we selected the brightest blue stars as candidates for the stellar spectroscopy. However, the brightest star observed by \citet{Gull:2022}, has a $B$ magnitude of 20.1 (in the photometric catalogue of \citealt{Stonkute:2014}), while our sample includes, in addition to the object in common mentioned above, seven blue ($B-V$ around $-0.2$ to 0.1) stars brighter than this limit. 

The outcome of our stellar classification procedure is reported in Column~(7) of Table~\ref{table:1}, and
a few notes are presented below:

\begin{enumerate}

\item star 02: the emission at \hbeta\ appears to be stellar since no extended nebular emission is detected at the wavelength of the Balmer lines and of \oii\lin3727. This is the brightest object in our sample. We assign an Iab luminosity class because, although
\ewhgamma\ is consistent with the Ia class, the \hgamma\ line is likely partially filled with emission. 

\item  in some cases \ewhgamma\ is considerably larger than what the \citet{Azzopardi:1987} calibration yields for the Ib luminosity class, but is intermediate between classes Ib and III in \citet{Balona:1974}. For these objects, we assigned the luminosity class II (01, 03, 10). For star 09 the \ewhgamma\ value we observe is consistent with class III in \citet{Balona:1974}. However, if we make a comparison with the absolute magnitudes of Galactic stars of the same spectral types we encounter a large discrepancy for the three A stars. The mean absolute magnitudes of A0 stars, calibrated with Hipparcos parallaxes by \citet{Wegner:2007}, are $M_V=-1.46$ (A0~II) and $M_V=-0.09$ (A0~III), \ie\ much fainter than observed for our targets (between approximately $-2.9$ and $-4.7$). Only with a luminosity classification of Ib we would find a correspondence with the Galactic stars.
In the case of the B5~II star in our sample (01) there is no such issue: we measure $M_V\simeq-3.0$, while for this classification \citet{Wegner:2006} gives $M_V=-2.82$.

This result could be hinting at the inadequacy of the method used here to estimate the luminosity class of our extremely metal-poor targets, at least of the A types. We have classified early-A stars as class II objects in other galaxies (\eg\ WLM and IC~1613: \citealt{Bresolin:2006a, Bresolin:2007a}), but their absolute magnitudes are in line with the expectations from Galactic equivalents. On the other hand, as explained in Sec.~\ref{sec:quantitative}, star 03 is found to be peculiar compared to the other stars, with a significantly fainter bolometric magnitude compared to the remaining stars.

\item star 08 is a O9~V star located inside \hii\ region SHK~1 (\citealt{Strobel:1991}). We are not aware of published spectra of this ionized nebula.  This is star K1/B1 of \citet{Gull:2022}.
Nebular contamination is clearly seen at \hbeta, as shown by the spectrum in Fig.~\ref{fig:spectra}, while the two-dimensional spectra also show weak extended emission at \hgamma.

\item the O7/O8~V star 12 is the likely ionizing source of \hii\ region SHK~4 (\citealt{Strobel:1991}), which is region $+112-020$ of \citet{van-Zee:2006}. The classification is made somewhat uncertain by the fact that the \hei\ lines, used to estimate the spectral type, are filled in by emission, with \lin4713 in emission and \lin4921 completely filled in.
We do not detect these lines in emission in the two-dimensional spectrum. Even \heii\lin4686 could be partially filled in. 
The fact that the \oiii\llin4959,\,5007 lines are easily seen in the two-dimensional spectrum is consistent with a higher ionization degree of SHK~4 compared with SHK~1, where such lines remain undetected. This is indirect evidence that star 12 is of an earlier spectral subtype than star 08.

\end{enumerate}

%==================================================================================
\section{Quantitative analysis} \label{sec:quantitative}
The goal of the quantitative spectral analysis is to determine the stellar atmospheric parameters effective temperature \teff, surface gravity \logg~and, most importantly, metallicity $[Z]$ (defined as $[Z] = \log Z/$\zsun, where \zsun~is number fraction of solar metallicity). Following the spectral classification we focus on the supergiants of spectral type late B and A and luminosity class Ib and II. Contrary to the O and early-B stars in our sample these objects have a relatively rich spectrum of metal lines and will allow for a reasonable metallicity determination even at the low metallicity of \leoa. Moreover, because of their relative brightness, the spectra have a high signal-to-noise ratio. This leaves us with seven objects for the quantitative analysis.

The analysis technique has been described in detail in previous papers (\citealt{Kudritzki:2014, Kudritzki:2016}, \citealt{Hosek:2014}, \citealt{Bresolin:2022}). We compare the normalized observed spectra with synthetic spectra from a comprehensive grid of metal line-blanketed model atmospheres with extensive non-LTE line formation calculations using elaborate atomic models (\citealt{Przybilla:2006}). The model grid described in \citet{Kudritzki:2008,Kudritzki:2012} is adopted, comprising effective temperatures from 7900\,K to 15000\,K and gravities between 0.8 to 3.0 dex in cgs units (the exact upper and lower limits depend on \teff, see Figure~1 in \citealt{Kudritzki:2008}). The metallicity $[Z]$ of the original grid ranges from $-1.30$ dex to $+0.50$ dex, however, because of the low metallicity of \leoa\ we have extended the grid to $[Z] = -1.45$ dex.

% - - - - - - - - - - - - - - - - - - - - - - - - - - - - - - - - - - - - - - - - - 
\begin{figure}
	\center \includegraphics[width=1\columnwidth]{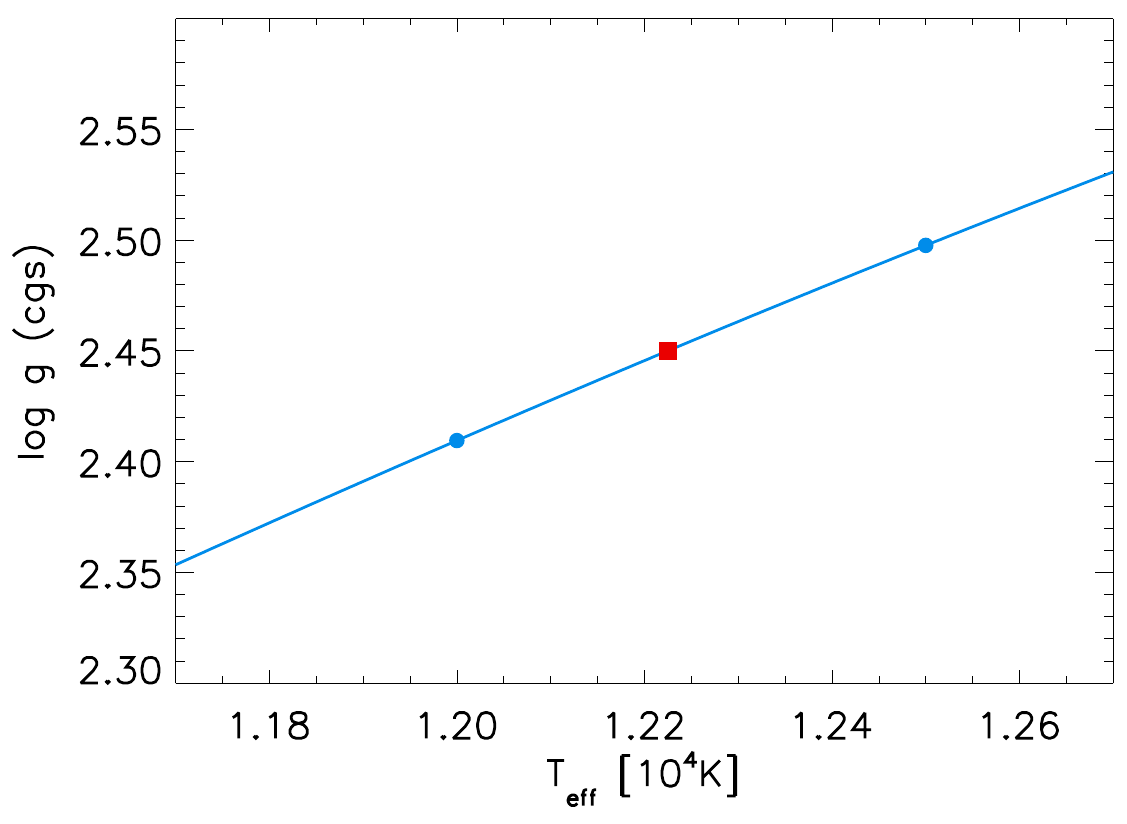}\medskip
	\caption{Balmer line fit of target 04 of Table 1. The blue curve shows the gravity \logg~at which the best fit of the Balmer lines is obtained for each value \teff\ on the abscissa. The red square shows the final \logg~and \teff\ values following from the additional fit of the metal lines along this curve (see text). The final \logg~and \teff\ for all targets analysed are given in Table 4.}\label{fig:Balmeriso}
\end{figure}
% - - - - - - - - - - - - - - - - - - - - - - - - - - - - - - - - - - - - - - - - - 

% - - - - - - - - - - - - - - - - - - - - - - - - - - - - - - - - - - - - - - - - - 
\begin{figure*}
	\center \includegraphics[width=0.7\textwidth]{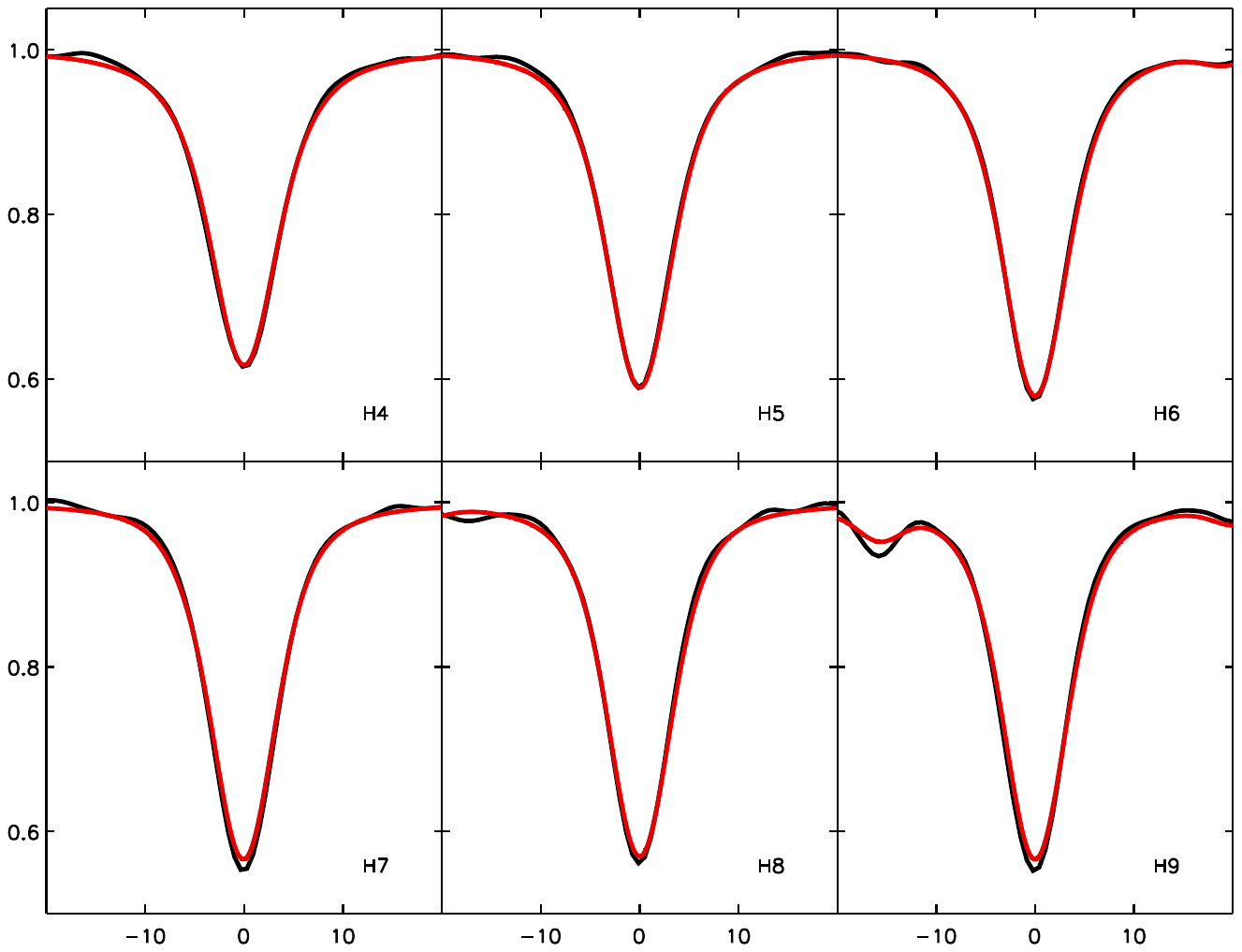}\medskip
	\caption{Balmer line fit of target 04 using the adopted \teff, \logg, and $[Z]$ values in Table 2. Comparison of model (red) and observed (black) Balmer line profiles. The abscissa is the displacement from the line center in \AA.}\label{fig:Balmerfit}
\end{figure*}
% - - - - - - - - - - - - - - - - - - - - - - - - - - - - - - - - - - - - - - - - - 

In a first step, we use the observed Balmer lines (H$_4$ to H$_{10}$) to constrain \logg~as a function of \teff~by finding the gravity at each fixed value of \teff\ that fits the Balmer lines best. Figure~\ref{fig:Balmeriso} gives an example for target 04 of Table~\ref{table:1}. The accuracy of the fit at each \teff~is 0.05 dex.  Figure~\ref{fig:Balmerfit} shows the fit of the Balmer line profiles for the same target 04. The quality of the fit is excellent and is typical of what we obtain for all the remaining targets, for which the fits are not shown.

\begin{figure}
%    \center \includegraphics[width=0.70\columnwidth]{f5a}\medskip
%    \center \includegraphics[width=0.70\columnwidth]{f5b}\medskip
%    \center \includegraphics[width=0.70\columnwidth]{f5c}\medskip
\gridline{\fig{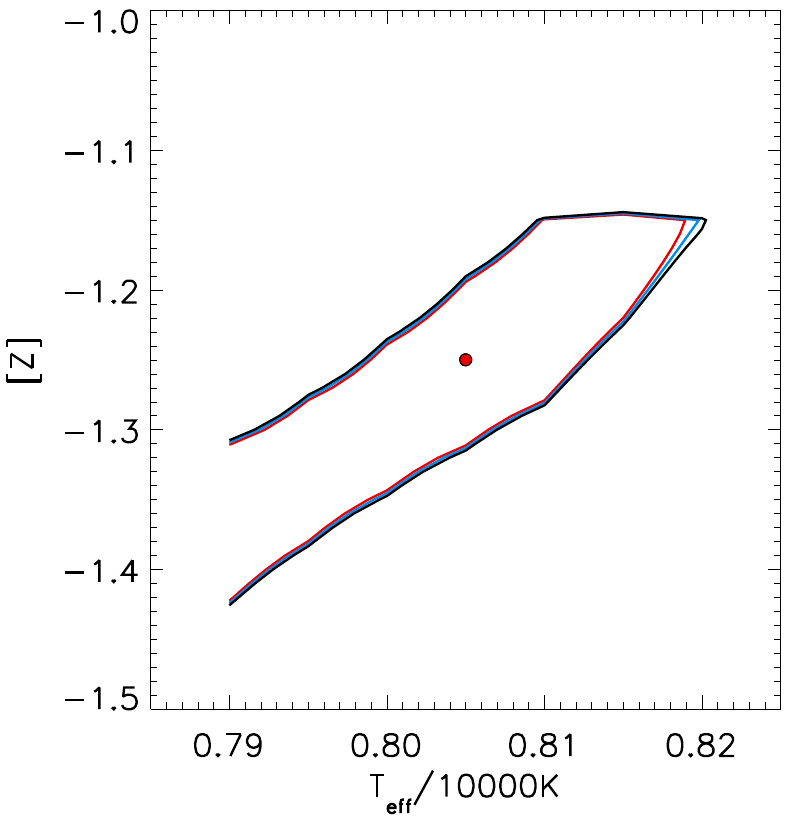}{0.35\columnwidth}{}}
\gridline{\fig{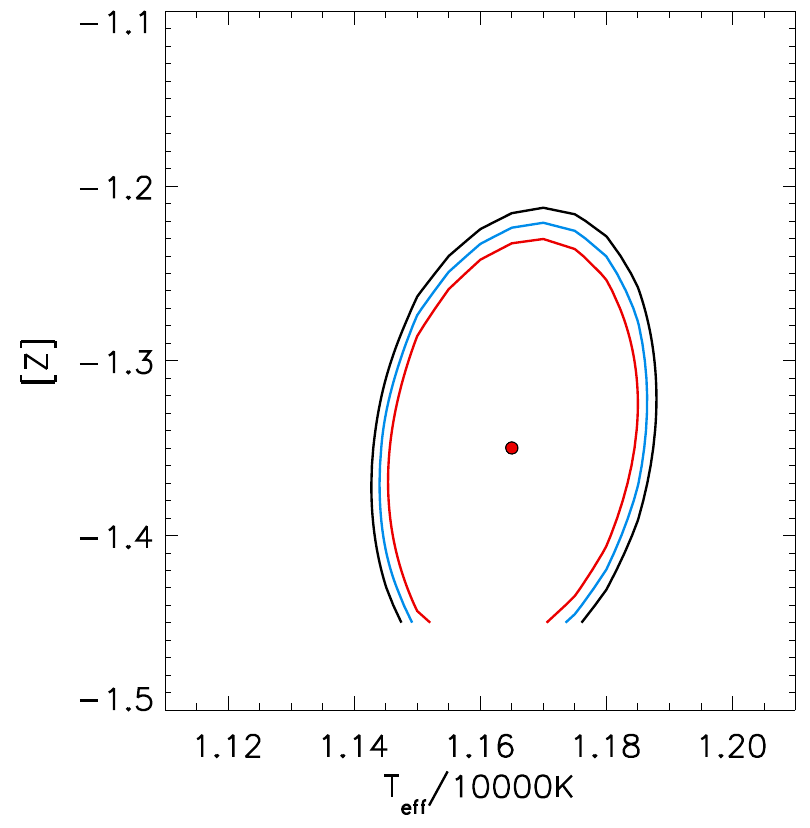}{0.35\columnwidth}{}}
\gridline{\fig{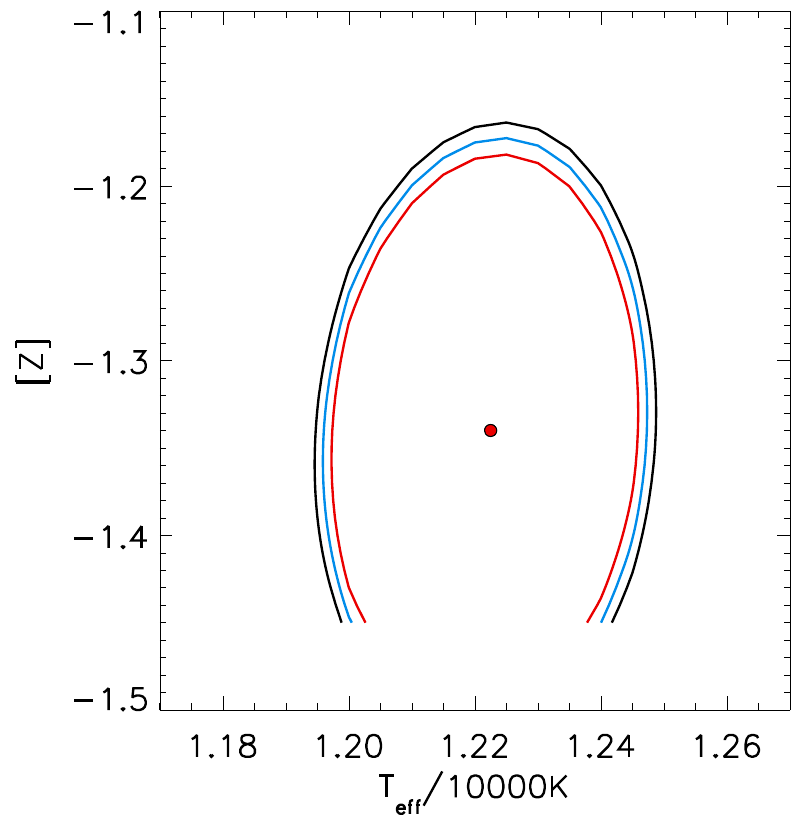}{0.35\columnwidth}{}}
	\caption{Determination of temperature and metallicity from $\Delta \chi^2$ isocontours obtained through the comparison of observed and synthetic spectra (see text). Plotted are $\Delta \chi^2$\,=\,3 (red), 6 (blue), and 9 (black) and the final fit values (red dot). The red and the black isocontours correspond to the 1- and 2-$\sigma$ uncertainties, respectively. {\em (Top):} target 10. {\em (Middle)} target 11. {\em (Bottom):} target 04.}\label{fig:isofit}
\end{figure}

\begin{figure}
%	\center \includegraphics[width=1\columnwidth]{f6a}\medskip
%    \center \includegraphics[width=1\columnwidth]{f6b}\medskip
\gridline{\fig{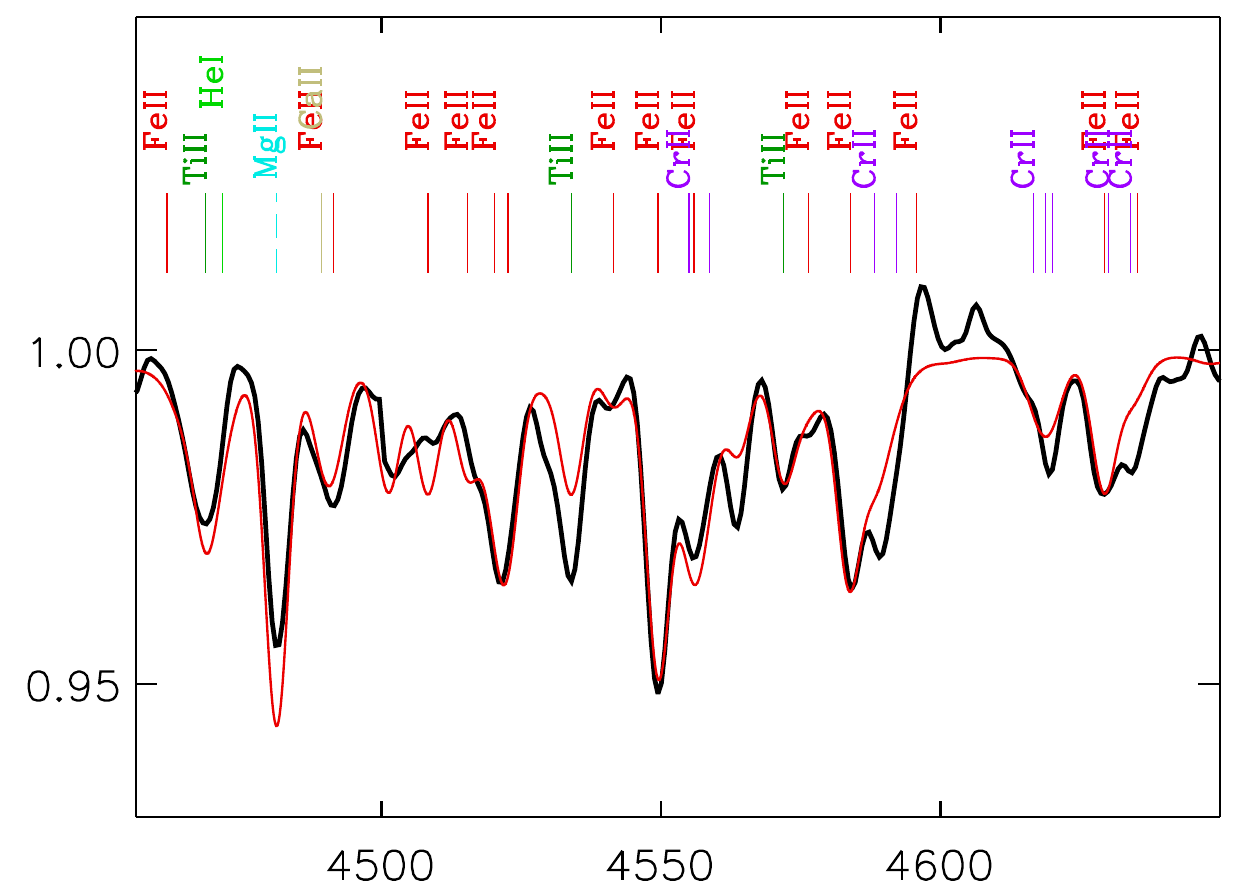}{0.7\columnwidth}{}}
\gridline{\fig{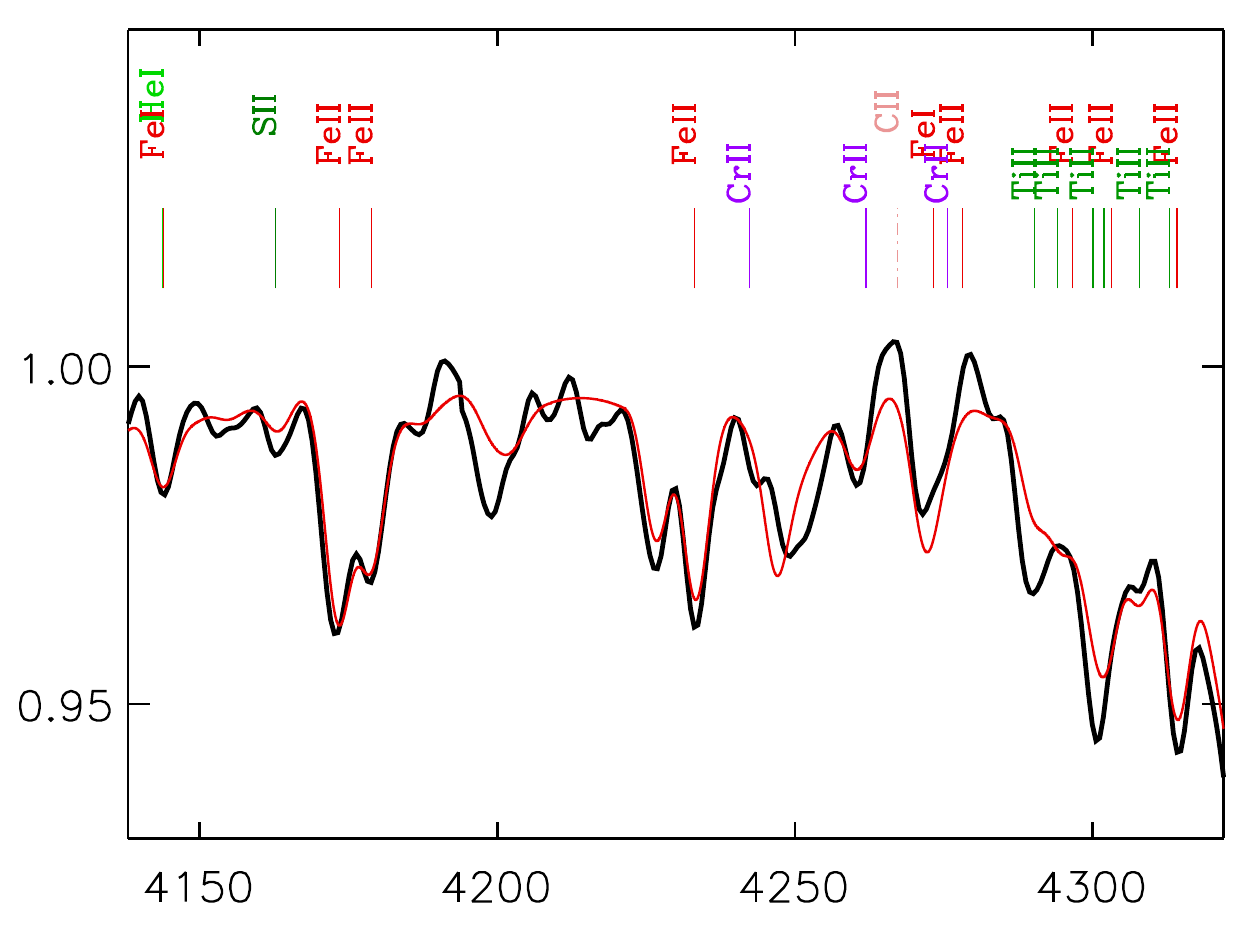}{0.7\columnwidth}{}}
	\caption{Target 10: observed metal lines (black) and fit with the synthetic spectrum calculated for the final stellar parameters (red) in selected parts of the spectrum.}\label{fig:metfit1}
\end{figure}

\begin{figure}
%	\center \includegraphics[width=1\columnwidth]{f7a}\medskip
%    \center \includegraphics[width=1\columnwidth]{f7b}\medskip
%    \center \includegraphics[width=1\columnwidth]{f7c}\medskip
\gridline{\fig{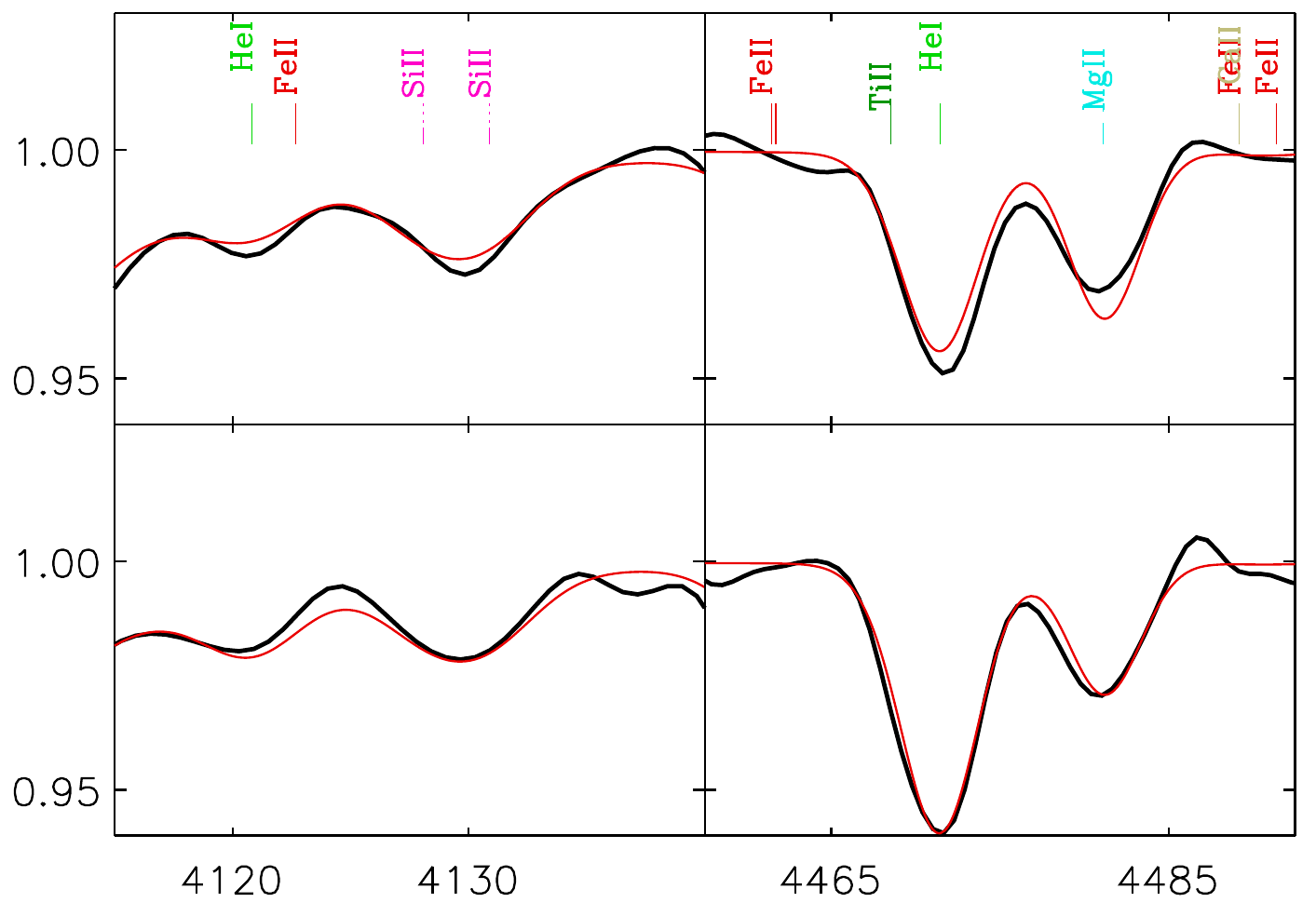}{0.4\columnwidth}{}}\medskip
\gridline{\fig{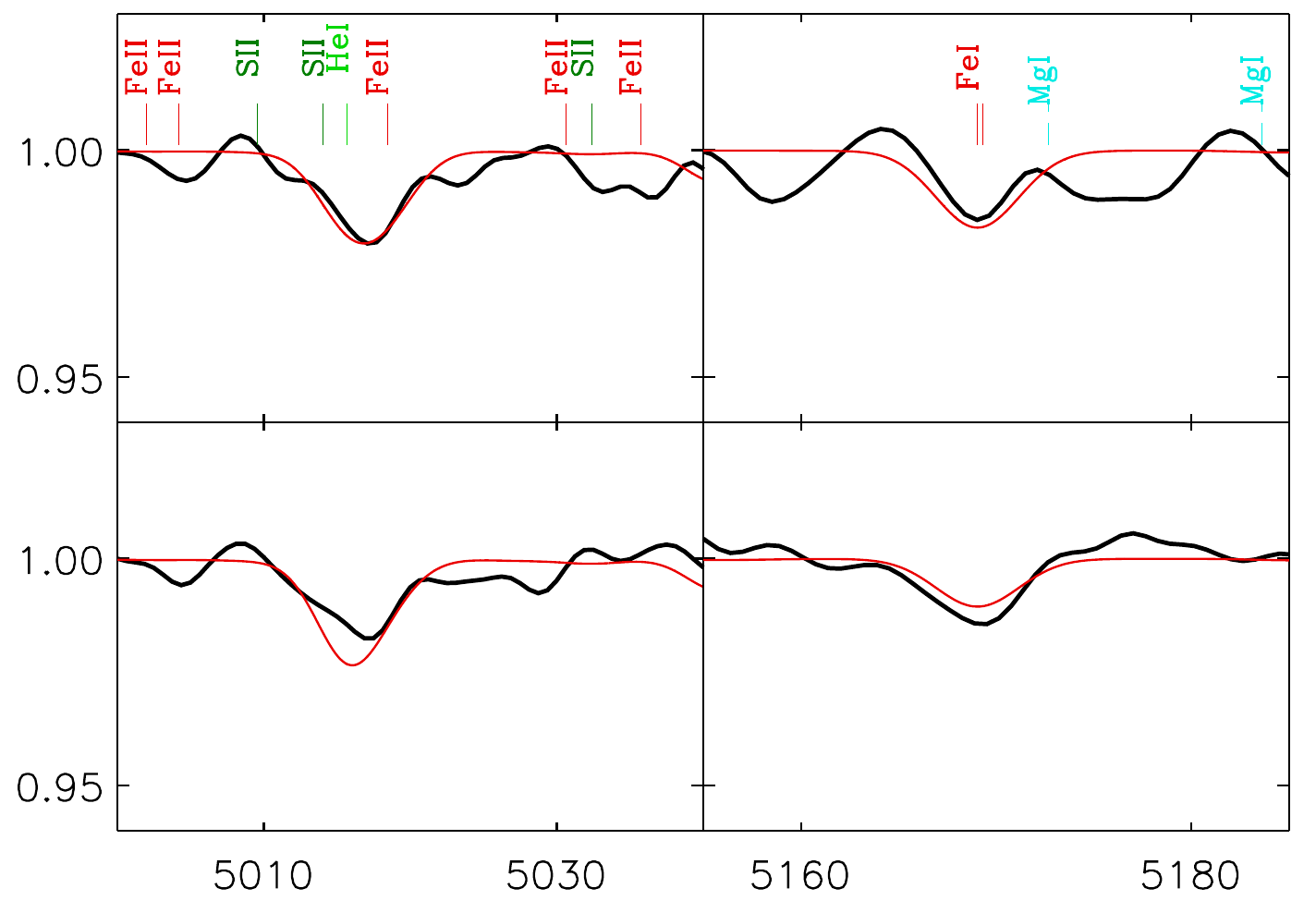}{0.4\columnwidth}{}}\medskip
\gridline{\fig{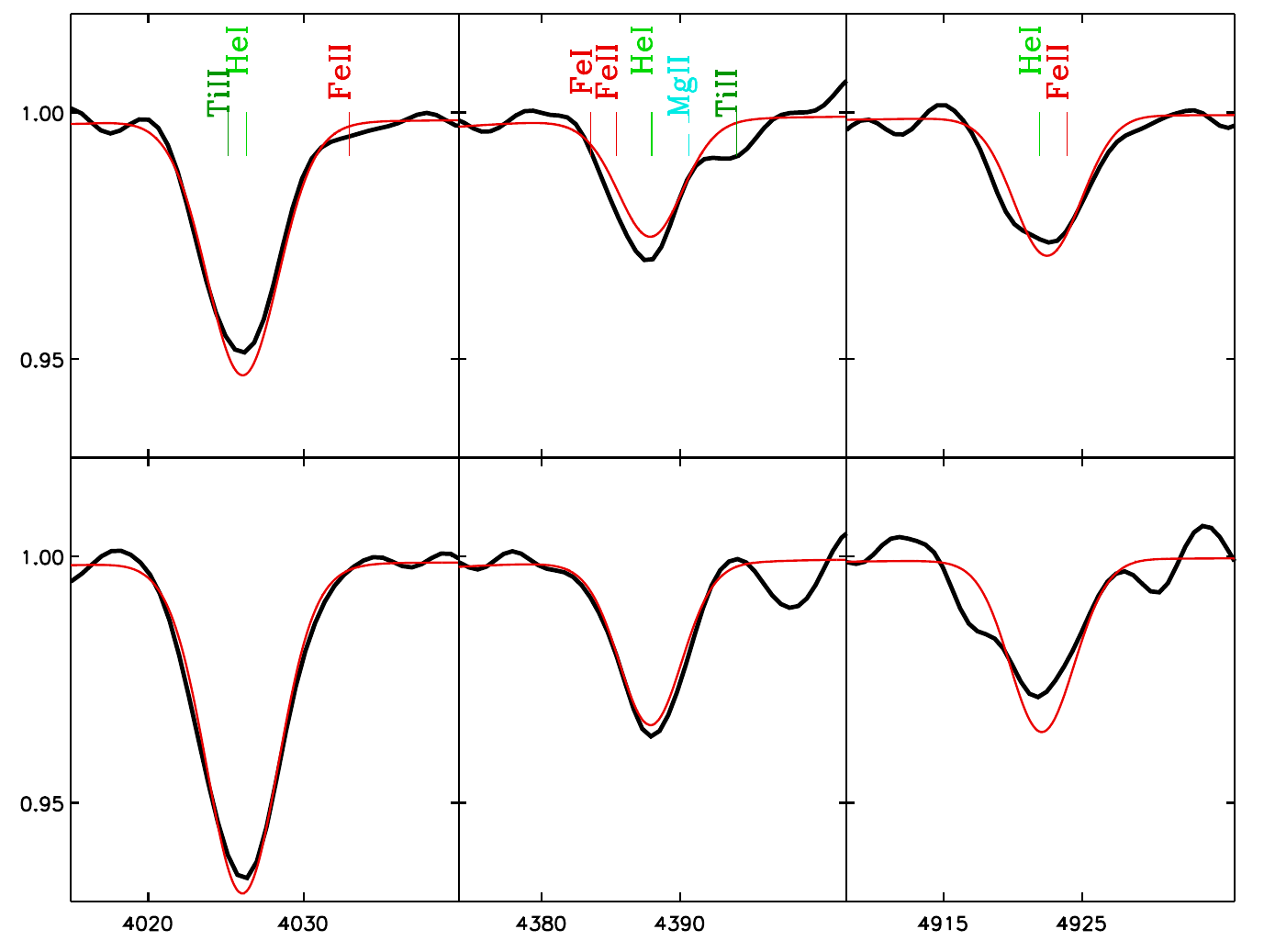}{0.4\columnwidth}{}}\medskip

 \caption{ Targets 11 (at the top in each panel) and 04 (bottom): observed metal lines (black) and fit with the synthetic spectrum calculated for the final stellar parameters (red) in selected parts of the spectrum.}\label{fig:metfit2}
\end{figure}

Subsequently, we move along the gravity-temperature relationship defined by the fit of the Balmer lines and compare observed and calculated fluxes as a function of metallicity in metal-line dominated spectral windows located away from the Balmer lines. We calculate $\chi^2$ values at each \teff~and $[Z]$. We determine the minimum of $\chi^2$ and isocontours in $\Delta \chi^2$ around the minimum, which provides us with the determination of \teff~and $[Z]$ and their uncertainties. From detailed Monte Carlo simulations (\citealt{Hosek:2014, Kudritzki:2013}) we know that the $\Delta \chi^2$ = 3 and 9 isocontours provide a conservative estimate of the 1- and 2-$\sigma$ uncertainties, respectively (the minimum $\chi^2$ values are of the order of unity). Figure~\ref{fig:isofit} displays examples for three of the targets and Figures~\ref{fig:metfit1}-\ref{fig:metfit2} show typical metal line fits.

% - - - - - - - - - - - - - - - - - - - - - - - - - - - - - - - - - - - - - - - - - 
%\begin{figure}
%\gridline{\fig{fig1a.pdf}{0.3\textwidth}{(a)}
%          \fig{fig1b.pdf}{0.3\textwidth}{(b)}
%          \fig{fig1c.pdf}{0.3\textwidth}{(c)}}
%\gridline{\fig{fig1d.pdf}{0.3\textwidth}{(d)}
%          \fig{fig1e.pdf}{0.3\textwidth}{(e)}}
%\gridline{\fig{fig1f.pdf}{0.3\textwidth}{(f)}}
%\caption{A nice inverted pyramid figure 
%consisting of six individual files}
%\end{figure}

% - - - - - - - - - - - - - - - - - - - - - - - - - - - - - - - - - - - - - - - - - 

%==================================================================================
%\floattable
\begin{deluxetable*}{lCCcccc}
	\tabletypesize{\footnotesize}	
	\tablecolumns{7}
	\tablewidth{0pt}
	\tablecaption{Stellar parameters.\label{table:2}}
	
	\tablehead{
		\colhead{ID}	     		&
		\colhead{\teff}	 			&
		\colhead{\logg}	 			&
		\colhead{$[Z]$}			&
		\colhead{\loggf}			&
		\colhead{$E(B-V)$}				&
		\colhead{m$_{bol}$}\\[-2.ex]
		\colhead{}       			&
		\colhead{(K)}       	&
		\colhead{(cgs)}       	&
		\colhead{(dex)}					& 	
		\colhead{(cgs)}					& 
		\colhead{(mag)}				& 														
		\colhead{(mag)}  } 	
	
	\colnumbers
	\startdata
	\\[-4.5ex]
	03 &  8450$\pm200$ & 1.81$\pm0.05$  & $-1.33\pm0.10$ & $2.11\pm0.15$ & $0.01^{+0.03}_{-0.01}$ & $21.09\pm0.11$ \\[-0.4ex] 
04 & 12225$\pm250$ & 2.45$\pm0.05$  & $-1.34\pm0.12$ & $2.10\pm0.06$ & $0.03\pm0.03$ & $18.78\pm0.12$ \\[-0.4ex] 
05 & 11850$\pm200$ & 2.63$\pm0.05$  & $-1.28\pm0.12$ & $2.34\pm0.06$ & $0.03\pm0.03$ & $19.47\pm0.11$ \\[-0.4ex] 
06 & 12725$\pm250$ & 2.54$\pm0.05$  & $-1.45\pm0.17$ & $2.12\pm0.06$ & $0.03\pm0.03$ & $18.87\pm0.12$ \\[-0.4ex] 
07 & 12275$\pm200$ & 2.24$\pm0.05$  & $-1.45\pm0.13$ & $1.88\pm0.05$ & $0.03\pm0.03$ & $18.46\pm0.11$ \\[-0.4ex] 
10 &  8050$\rm^{+150}_{-250}$ & 1.80$\pm0.05$ & $-1.25^{+0.10}_{-0.25}$ & $2.18^{+0.25}_{-0.15}$ & $0.10\pm0.03$ & $19.40\pm0.11$ \\[-0.4ex] 
11 & 11650$\pm200$ & 2.58$\pm0.05$  & $-1.35\pm0.12$ & $2.32\pm0.06$ & $0.06\pm0.03$  & $19.16\pm0.11$ \\[-0.4ex]   
	\\[-2.5ex]
	\enddata
	\tablecomments{The \logg\ uncertainty is for a fixed \teff\ value. }
\end{deluxetable*}

%==================================================================================

We note from Figure~\ref{fig:isofit} that some of our targets have stellar parameters which place them at the edges of our model grid. While its best fit occurs for \teff~= 8050\,K, target 10 could have a temperature below 7900\,K, which is the lower temperature limit of our grid. In this case, the metallicity would be smaller than $[Z] = -1.40$. This is the only star at the temperature edge of our grid.

Targets 10 and 11 in Figures~\ref{fig:metfit1}-\ref{fig:metfit2} have metallicities at the lower edge of our grid but the shape of the isocontours allows an estimate of the uncertainty towards lower $[Z]$ values. We encounter a similar situation for targets 06 and 07, whereas targets 03 and 05 have 1-$\sigma$ isocontours located fully inside the grid plane.

The spectroscopic parameters of the seven targets analyzed are given in Table~\ref{table:2}. All have very low metallicities, with values between $[Z]= -1.25$ and $-1.45$. The average metallicity value is $\bar{[Z]} = -1.35\pm0.08$. We will discuss this result below in relation to all the other star-forming galaxies studied so far in our project. We note that our average stellar metallicity is in close agreement with results from the analysis of \hii\ region emission lines: $[Z] = -1.31\pm0.1$ (\citealt[semi-empirical method]{van-Zee:2006}) and $[Z] = -1.3\pm0.2$ (\citealt[combination of direct method and photoionization modelling]{Ruiz-Escobedo:2018}). The direct oxygen abundance based on the detection of the \oiii\lin4363 line in a single \hii\ region yields
$[Z] = -1.30\pm0.10$ (\citealt{Ruiz-Escobedo:2018}).

% - - - - - - - - - - - - - - - - - - - - - - - - - - - - - - - - - - - - - - - - - 
% - - - - - - - - - - - - - - - - - - - - - - - - - - - - - - - - - - - - - - - - - 

% - - - - - - - - - - - - - - - - - - - - - - - - - - - - - - - - - - - - - - - - - 
% - - - - - - - - - - - - - - - - - - - - - - - - - - - - - - - - - - - - - - - - - 

% - - - - - - - - - - - - - - - - - - - - - - - - - - - - - - - - - - - - - - - - - 
\begin{figure}
	\center \includegraphics[width=1\columnwidth]{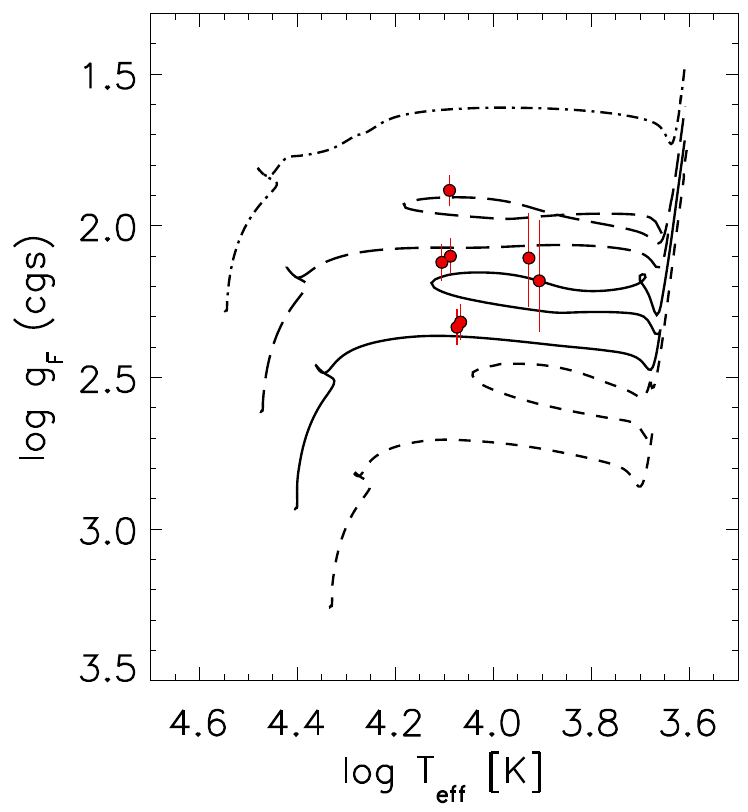}\medskip
	\caption{Spectroscopic Hertzsprung-Russel diagram of blue supergiant stars in \leoa\ together with MESA stellar evolution tracks calculated for $[Z] = -$1.25 (\citealt{Choi:2016, Dotter:2016}). The tracks include the effects of stellar rotation and are calculated for initial main-sequence masses of 5\,\msun\ (dashed), 7\,\msun\ (solid), 10\,\msun\ (long-dashed) and 15\,\msun\ (dashed-dotted).}\label{fig:sHRD}
\end{figure}
% - - - - - - - - - - - - - - - - - - - - - - - - - - - - - - - - - - - - - - - - - 

Table~\ref{table:2} also contains the values of the flux-weighted gravity \loggf~= $\logg - 4\log$(\teff/10$^4$\,K) of each target. As found by \citet{Kudritzki:2003}, \loggf~can be used to determine accurate spectroscopic distances through the flux-weighted gravity--luminosity relationship (\fglr) and we will apply this technique in Sect.~\ref{sec:distance}. In addition, the value of \loggf~allows an assessment of evolutionary status and stellar masses through the spectroscopic Hertzsprung-Russel diagram (sHRD, see \citealt{Langer:2014}), which is distance independent. Figure~\ref{fig:sHRD} presents this diagram for our \leoa\ \bsg\ targets, together with MESA stellar evolution tracks calculated for $[Z] = -$1.25 (\citealt{Choi:2016, Dotter:2016}). We infer masses for our target stars in the range 7 to 12 \msun (we note that $[Z] = -$1.25 is somewhat higher than our average metallicity value, but using the next lower value $-1.50$ in the MESA grid leads to very similar tracks and the same masses).  These masses are lower than those in the range 15-40 \msun~ found in our blue supergiant work in other galaxies (see \citealt{Kudritzki:2016}, \citealt{Bresolin:2022} and references therein) and the \loggf~values are correspondingly larger. This proceeds from the combination of the low stellar mass of this dwarf galaxy and the shape of the stellar initial mass function, which makes the presence of more massive stars more unlikely.

We note that all our seven spectroscopic targets are located in a region of the sHRD where stellar evolution theory predicts blue loop excursions for stars climbing the red giant branch. This is a consequence of the masses of our targets and the low metallicity. At larger masses and higher metallicity ($[Z] \ge -0.90$) the MESA tracks do not show blue loops in the temperature range 
between 8000\,K and 13000\,K of our observed objects. Using the evolutionary tracks and focusing on this temperature range to estimate the time spent in the first HRD crossing towards the red giant branch and in the 
blue loops we find that the latter is larger by a factor of 30 to 120.
Thus, it is extremely likely that our objects are stars in the blue loop stage. Note, however, our remarks concerning target 03 at the end of this section.

All stars included in this section on the quantitative analysis, except object 10, are identified as blue loop stars by \citet{Lescinskaite:2022} based on their position in color-magnitude diagrams. With the aid of stellar isochrones they derived ages between 30~Myr and 138~Myr for the targets we have in common.

Using the spectroscopic stellar parameters of \teff, \logg, and $[Z]$ we calculate model spectral energy distributions (SEDs) and colors $B-V$ and $V-I$. The comparison with the observed colors of Table~\ref{table:1} yields interstellar absorption color excesses $E(B-V)$ and $E(V-I)$. With the relation $E(B-V) = 0.75\,E(V-I)$ (derived by applying
the \citealt{Cardelli:1989} reddening law to
the model SEDs) we obtain two $E(B-V)$ values for each object and calculate the mean, which is given in Table~\ref{table:2}. We then compute the interstellar extinction $A_V = R_{V} \cdot E(B-V)$ with $R_V = 3.3$ to correct the $V$ magnitudes for reddening and apply the model atmosphere bolometric corrections to obtain apparent bolometric magnitudes \mbol. The latter are included in Table~\ref{table:2}. We note that the amount of reddening is very small. This may be a selection effect because bright and blue objects were selected from the color magnitude diagrams for our spectroscopy.

Inspection of Table \ref{table:2} immediately reveals that target 03 is more than two magnitudes fainter than all the other targets of comparable \loggf. It is by far the faintest object in our sample, while it should be among the brightest according to its flux-weighted gravity. We will discuss this target in detail in Sec.~\ref{sec:postagb}.

%==================================================================================
\section{Leo A and the mass--metallicity relationship of star-forming galaxies} \label{sec:MZR}

\begin{figure*}
	\center \includegraphics[width=0.8\textwidth]{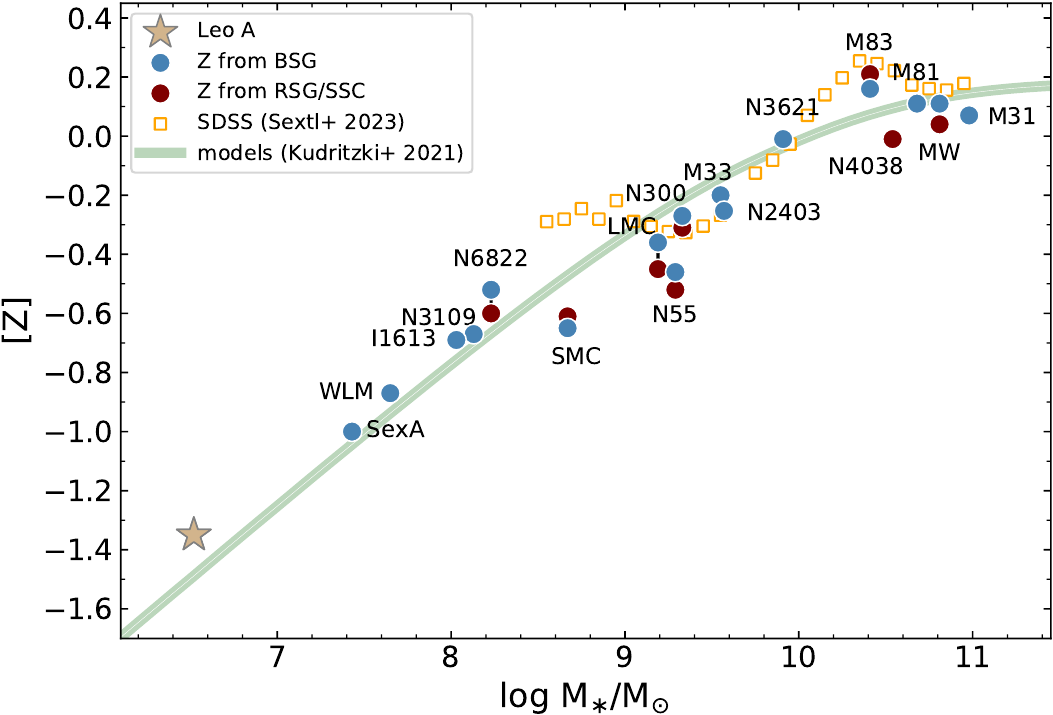}\medskip
	\caption{The mass-metallicity relationship of star-forming galaxies based on absorption line studies of the young stellar population. Results from blue supergiants (BSG) are shown in blue, while metallicities from red supergiants (RSG) and superstar clusters (SSC) are displayed in red. The yellow symbols correspond to the recent population synthesis study of 250,000 star-forming SDSS galaxies by \citet{Sextl:2023}. Our new result obtained for \leoa\ is represented with the star symbol. Predictions from the galaxy evolution look-back models by \citet{Kudritzki:2021} are shown as the green curve. }\label{fig:MZR}
\end{figure*}

With a stellar mass M$_* = 3.3\times10^6$\,\msun\ (\citealt[corrected for the different galaxy distance adopted here]{Leroy:2019}) \leoa\ is the galaxy that has by far the lowest mass of the systems studied to date in our project (see \citealt{Bresolin:2022} for a summary). It is, thus, ideal to investigate how the tight relationship between stellar mass and stellar metallicity of the young stellar population (mass--metallicity relation, \mzr) extends to very low galaxy stellar masses. Figure~\ref{fig:MZR} shows how \leoa\ marks the end of a tight relationship stretching over four orders of magnitude in stellar mass. This confirms that the \mzr\ can be nicely explained by simple galaxy evolution models that assume that galactic winds and matter infall result in a simple redshift-dependent relationship between galaxy gas mass and stellar mass (\citealt{Kudritzki:2021}). In this sense \leoa\ provides a crucial contribution towards the understanding of galaxy evolution. 

In Figure~\ref{fig:MZR} we have also added the results of the most recent study by \citet{Sextl:2023}. In this work spectra of 200000 SDSS star-forming galaxies stacked in bins of galaxy stellar mass were analysed using a population synthesis technique which allows to disentangle metallicity and ages of the young ($\sim$ 0.2 Gyr) and old ($\sim$ 10 Gyr) stellar population. The metallicities of the young population in the different galactic stellar mass bins are shown in Figure 9 as orange squares. They are in good agreement with metallicities obtained from the analysis of \bsg s in nearby galaxies, confirming the concept of a universal mass-metallicity relationship in the low-redshift universe.

\section{Flux-weighted gravity--luminosity relationship and distance estimate} \label{sec:distance}

% - - - - - - - - - - - - - - - - - - - - - - - - - - - - - - - - - - - - - - - - - 
\begin{figure}
	\center \includegraphics[width=1\columnwidth]{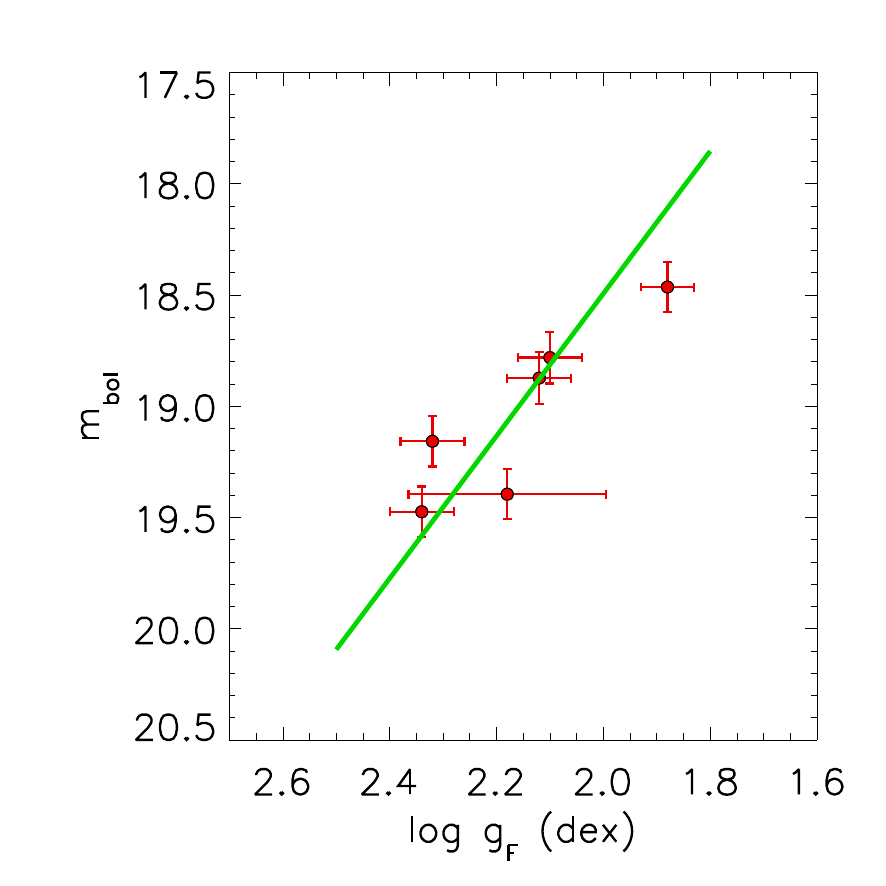}\medskip
	\caption{\fglr\ of \leoa\ blue supergiant stars. The green line is the fitting relation described in the text.} \label{fig:FGLR}
\end{figure}
% - - - - - - - - - - - - - - - - - - - - - - - - - - - - - - - - - - - - - - - - - 

% - - - - - - - - - - - - - - - - - - - - - - - - - - - - - - - - - - - - - - - - - 
\begin{figure}
	\center \includegraphics[width=1\columnwidth]{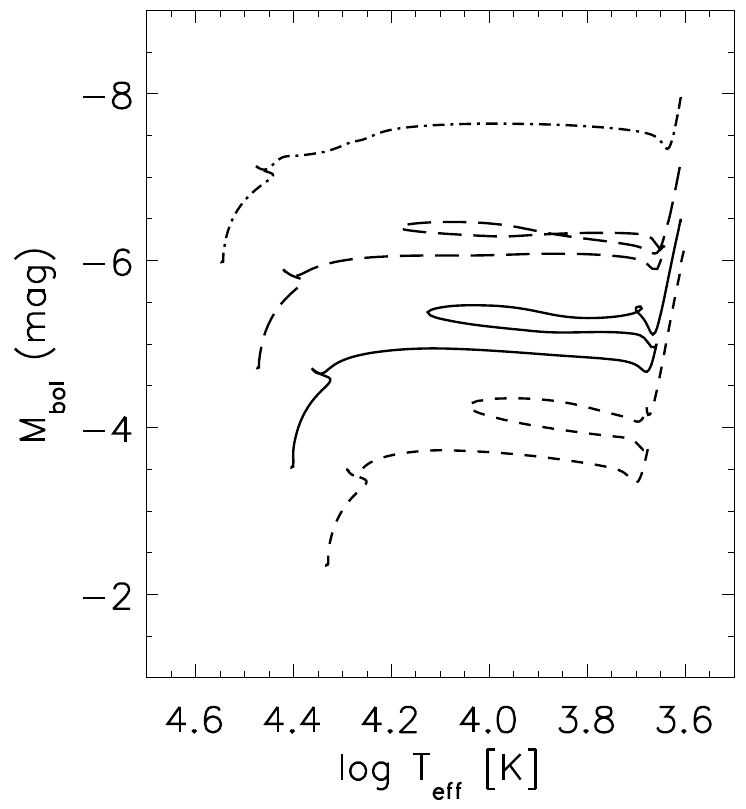}\medskip
	\caption{Hertzsprung-Russel diagram of MESA stellar evolution tracks calculated for $[Z] = -1.25$ (\citealt{Choi:2016, Dotter:2016}). The tracks include the effects of stellar rotation and are calculated for initial main-sequence masses of 5\,\msun\ (dashed), 7\,\msun\ (solid), 10\,\msun\ (long-dashed) and 15\,\msun\ (dashed-dotted).}\label{fig:HRD}
\end{figure}
% - - - - - - - - - - - - - - - - - - - - - - - - - - - - - - - - - - - - - - - - - 

% - - - - - - - - - - - - - - - - - - - - - - - - - - - - - - - - - - - - - - - - - 
\begin{figure}
	\center \includegraphics[width=1\columnwidth]{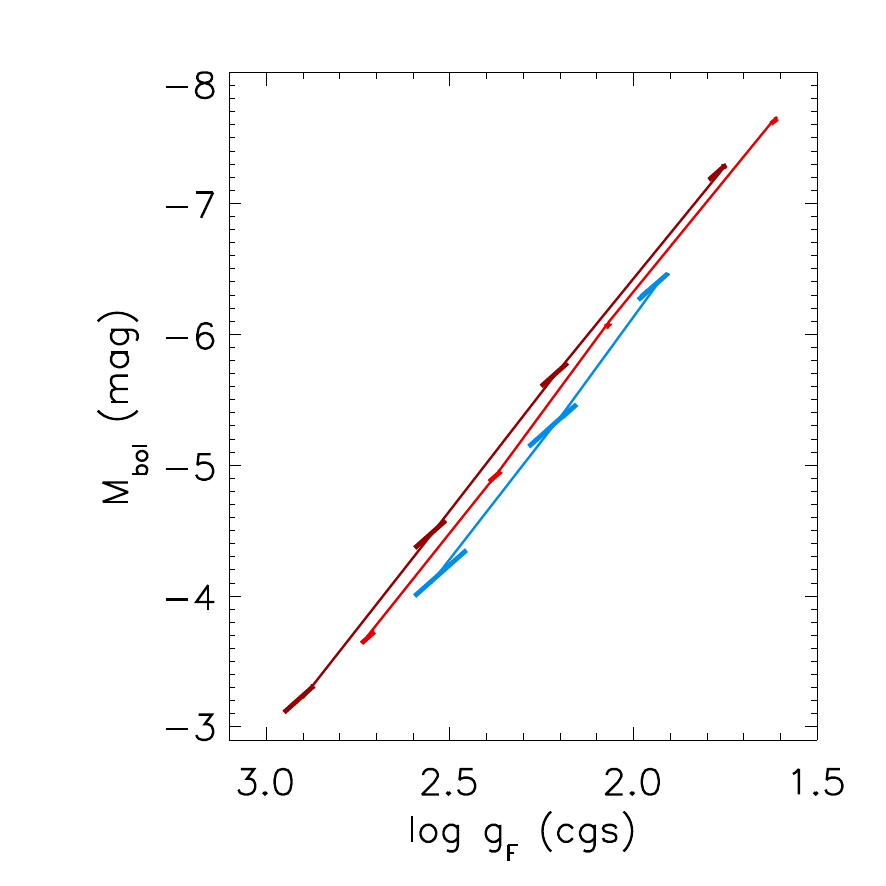}\medskip
	\caption{Stellar evolution \fglr\ at $[Z]=-1.25$ constructed from Figures~\ref{fig:sHRD} and \ref{fig:HRD} for \bsg s in the first crossing phase towards the red giant phase (red) and during the blue loop phase (blue). The \fglr\ plotted in dark red, also obtained from MESA evolutionary tracks, refers to the first crossing phase at $[Z] = -0.25$.}     \label{fig:evofglr}
\end{figure}
% - - - - - - - - - - - - - - - - - - - - - - - - - - - - - - - - - - - - - - - - - 

As found by \citet{Kudritzki:2003,Kudritzki:2008} blue supergiant stars exhibit a correlation between bolometric magnitude and flux-weighted gravity, the flux-weighted gravity--luminosity relationship (\fglr), which can be used to determine accurate distances to galaxies. Figure~\ref{fig:FGLR} shows this relationship for the \leoa\ \bsg s included in Table~\ref{table:2}. The figure does not contain object 03 of Table~\ref{table:2} because it is a faint outlier as already discussed above.

We adopt the calibration of the \fglr~by \citet{Urbaneja:2017}, obtained from the analysis of 90 \bsg s, adjusted to the LMC distance modulus $m-M = 18.477$ mag \citep{Pietrzynski:2019}:

\begin{equation}\label{eq:fglr1}
M_{bol} = 3.20\, (\log\,g_{\mbox{\tiny\it F}} - 1.5) - 7.878
\end{equation}

\noindent
and determine a \fglr-based distance modulus of \leoa\ from a regression fit accounting for the errors in \loggf~and \mbol. We obtain $m-M = 24.77\pm0.11$ mag. The green line in Figure~\ref{fig:FGLR} shows the fitting relation of Eq.~(\ref{eq:fglr1}), shifted by an amount equal to this distance modulus.

Our \fglr\ distance modulus is significantly larger than the value of $24.40\pm0.06$ determined by \citet{Nagarajan:2022} from RR Lyrae stars using a new Gaia eDR3-based calibration. While the ad hoc adoption of an RR Lyrae metallicity of $[Z] = -2.0$ may add some additional uncertainty to this value, the distance modulus of the metal-poor dwarf galaxy IC~1613 that these authors report, $m-M=24.42\pm0.03$ mag, is in excellent agreement with the \fglr\ value of $24.47\pm0.11$ mag (see \citealt{Bresolin:2022}). This seems to rule out large systematic effects. On the other hand, we also note that \citet{Nagarajan:2022} provide an average distance modulus literature value of $24.48\pm0.12$ for Leo A, which comes marginally close to the \fglr~value, if the errors are considered. The Extragalactic Distance Database \citep{Tully:2009} gives a tip of the red giant branch (\trgb) distance modulus of $24.35^{+0.47}_{-0.14}$ mag.

At face value the \fglr\ severely overestimates the distance to \leoa. However, there are a few important issues that must be considered to understand the reasons for the discrepancy. First of all, the sample of LMC \bsg s used by \citet{Urbaneja:2017} to obtain the calibration relationship of Eq.~(\ref{eq:fglr1}) consisted of objects of lower \loggf\ and higher \Mbol. No object in the LMC had a \loggf\ value larger than 1.90. The application of Eq.~(\ref{eq:fglr1}) for the distance determination is, thus, an extrapolation of the calibration relationship. The uncertainty introduced by the extrapolation is very likely small. The \bsg s in the low-metallicity dwarf galaxy WLM studied by \citet{Urbaneja:2008} cover the usual range between 1.1 and 1.9 in \loggf, but also include two objects with \loggf\,=\,2.2 and \loggf\,=\,2.3. The fit with our present \fglr\ calibration to the WLM supergiants yields good agreement with the \trgb\ distance (see \citealt{Sextl:2021}) and the two objects with high \loggf\ agree with the extrapolated relationship.

More importantly, as discussed above all of the objects used for the \fglr\ distance fit are very likely objects in the blue loop stage. This means that their \loggf\ values are about 0.2 dex lower than for \bsg s in the same temperature range crossing the HRD for the first time towards the red giant stage. These blue loop objects are brighter than they were
during their first crossing, but only by about 0.1 mag (see Figure~\ref{fig:HRD}). This implies that the \fglr\ of very low metallicity \bsg s in the mass range we are considering and captured in the blue loop stage will be fainter than the \fglr\ of objects captured in their first crossing. This is shown in Figure~\ref{fig:evofglr}, where we have connected the \loggf\ and \Mbol\ values of the MESA evolutionary tracks of Figures~\ref{fig:sHRD} and \ref{fig:HRD} in the \teff\ range 7900-13000\,K in order to construct the predicted \fglr\ for both first crossing and blue loop \bsg s. We see that the differential shift of the blue loop \fglr~amounts to $\sim$\,0.25 mag.

The third aspect in our \leoa\ \fglr\ discussion concerns the extremely low metallicity of the stars. The LMC \bsg\ calibration sample of \citet{Urbaneja:2017} had a metallicity of $[Z] = -0.35$. At the stellar luminosity and mass values of our \leoa\ targets there is an additional small metallicity effect, which impacts on both flux-weighted gravities and bolometric magnitudes. As Figure~\ref{fig:evofglr} illustrates,
the \fglr\ of the first crossing at the higher metallicity of $[Z]=-0.25$ (dark red curve) is 0.1 magnitudes brighter than at $[Z]=-1.25$ (red curve). It is worth pointing out that \citet{Meynet:2015} have already studied stellar evolution metallicity effects on the \fglr\ and found them to be small, but their work focused on higher stellar luminosities and did not consider the extremely low metallicities encountered in \leoa.

The combined differential effect of blue loops and extremely low metallicities amounts to a decrease of the bolometric magnitudes by 0.35 mag. The fact that we are using a \fglr\ calibration based on objects that are experiencing the first crossing of the HRD and approximately 10$\times$ the metallicities may explain why we obtain a distance modulus that is $\sim$0.35 mag larger than the one obtained by the careful RR Lyrae analysis of \citet{Nagarajan:2022}.

%==================================================================================
\section{The post-AGB object 03} \label{sec:postagb}

As we have already pointed out, object 03 of Table~\ref{table:2} is a faint outlier by at least 2 magnitudes in the \fglr. Assuming a distance modulus of 24.40 mag we obtain an absolute magnitude \Mbol\,=\,$-3.3$ mag for this star, corresponding to a luminosity $\log L/L_{\odot} = 3.2\pm0.1$, where the uncertainty comes from the \mbol\ error and a distance modulus error of 0.1 mag. This puts object 03 into the luminosity range of low-mass post-AGB objects, which have left the asymptotic giant branch and are on their way to becoming planetary nebulae. A typical low-metallicity post-AGB star with a mass of  0.52~\msun\ would have a luminosity in this ballpark (see, for instance, 
\citealt{Miller-Bertolami:2016}). Since \loggf\,=\,$\log M/M_{\odot} - \log L/L_{\odot} + 5.392$ (see \citealt{Langer:2014}) we obtain \loggf\,=\,$1.91\pm0.1$ dex for such an object, which agrees within the error margins with the flux-weighted gravity of object 03. We therefore conclude that target 03 is a low-mass post-AGB star. 

As a low-mass post-AGB star object 03 is significantly older than the \bsg s. Its age is very likely in the range of Gyrs. The fact that the metallicity we find, $[Z] = -1.33\pm0.10$, is similar to that of the \bsg s confirms that the chemical enrichment in \leoa\ has been proceeding very slowly (\citealt{Cole:2007, Kirby:2017}). In this context we also point out that the metallicity of object 03 is very similar to that of the single planetary nebula known in \leoa\ (\citealt{Skillman:1989}) and analyzed by \citet{van-Zee:2006}, who obtained $[Z] = -1.39\pm0.05$ (the re-analysis of the same data by \citealt{Ruiz-Escobedo:2018} gave $[Z] = -1.35\pm0.02$).

%==================================================================================
\section{Summary} \label{sec:summary}
In this paper, we have presented new spectroscopic data for 12 among the visually brightest blue stars in the dwarf irregular galaxy \leoa and carried out the first quantitative analysis of blue supergiants in this galaxy. Within our series of studies on the quantitative spectroscopy of evolved massive stars in nearby star-forming galaxies, this look at \leoa\ is of special significance, given the known low gas-phase metallicity of this system,  around 5\% of the solar value.

The classification of the spectra shows that all our targets are individual stars in \leoa. The two located within the boundaries of known \hii\ regions are late-O main-sequence stars and are likely powering the ionization of their respective host nebulae. The remaining objects are BA-type bright giants or supergiant stars, intrinsically less luminous than the majority of the stars we have investigated in other, more massive galaxies. We identify one of our targets as a post-AGB star, due to its significantly low bolometric luminosity for its value of the flux-weighted gravity, in comparison with the other stars that we have analyzed. The fact that the metallicity of this object is similar to that of the supergiant stars confirms that the chemical enrichment in \leoa\ is progressing very slowly.

We have measured the surface stellar parameters, \teff\ and \logg, as well as the metallicity of 
seven stars, all of the spectral types between B8 and A0, by comparing the rectified observed spectra with a large grid of model spectra, despite the weakness of the metal lines which could hamper the heavy element analysis. We obtain remarkably similar values of the metallicity for the seven stars, with an average $[Z] = -1.35\pm0.08$, in superb agreement with the published gas-phase chemical abundance obtained from \hii\ regions and one planetary nebula. This measurement allows us to place \leoa\ into the stellar mass--metallicity relation of star-forming galaxies derived from the spectroscopy of massive stars, rather than the emission-line analysis of the ionized gas, and which extends now over four orders of magnitude in mass, in nice agreement with theoretical predictions from simple galaxy evolution models.

From the stellar parameters and the flux-weighted--luminosity relation (\fglr) we derive a spectroscopic distance modulus $m-M = 24.77 \pm 0.11$. This result is significantly larger (by 0.37 mag) than the value indicated by RR Lyrae. This discrepancy is explained by the very low metallicity of \leoa, one order of magnitude below that of the \fglr\ calibrator (the LMC), and, most of all, by its effect on stellar evolution. In the \teff\ interval that characterizes our targets, stellar evolution models calculated at a metallicity comparable to the one measured in \leoa\ show that: {\em (i)} the objects are very likely in the blue loop phase and {\em (ii)} the \fglr\ is affected by the evolutionary stage of the stars as they evolve across the HRD. In fact, the relation is systematically fainter for stars that are observed while in the blue loop phase compared to those that are crossing the HRD for the first time towards the red giant phase.
Such an effect is absent at the higher metallicities found in the other galaxies we have analyzed.

%==================================================================================
\begin{acknowledgments}
This research has made use of the Keck Observatory Archive (KOA), which is operated by the W. M. Keck Observatory and the NASA Exoplanet Science Institute (NExScI), under contract with the National Aeronautics and Space Administration.
RPK acknowledges support by the Munich Excellence Cluster Origins funded by the Deutsche
Forschungsgemeinschaft (DFG, German Research Foundation) under Germany's Excellence Strategy EXC-2094 390783311.
The authors wish to recognize and acknowledge the very significant cultural role and reverence that the summit of Maunakea has always had within the indigenous Hawaiian community. We are most fortunate to have the opportunity to conduct observations from this mountain.
\end{acknowledgments}

\facility{Keck:I (LRIS)}

\software{IRAF (\citealt{Tody:1986, Tody:1993}), SciPy (\citealt{Virtanen:2020}), NumPy (\citealt{Harris:2020}), Matplotlib (\citealt{Hunter:2007a}), PyRAF (\citealt{Science-Software-Branch-at-STScI:2012}).}

%==========================================================================================================

\bibliographystyle{aasjournal}
\bibliography{Papers}
%\bibliography{References}

\end{document}